\documentclass{article}
\usepackage{amssymb}

\begin{document}

\title{{\bf Quantized Detector Networks: a Quantum Informational
Approach to the Description and Interpretation of
Quantum Physics}} \author{{\bf George Jaroszkiewicz}\\
School of Mathematical Sciences, University of Nottingham, \\
University Park, Nottingham NG7 2RD, UK}
\maketitle

\begin{abstract}
We discuss the QDN (quantized detector network) approach to the formulation
and interpretation of quantum mechanics. This approach gives us a
system-free approach to quantum physics. By this, we mean having a proper
emphasis on those aspects of physics which are observable and an avoidance
of metaphysical concepts, which by definition are incapable of verification
and should play no role in science on that account. By focusing on only what
experimentalists deal with, i.e., quantum information, we avoid the
ambiguities and confusion generated by the undue objectification of what are
complex quantum processes.
\end{abstract}

\section{Introduction}

The conceptual problems which arise in the interpretation of quantum
mechanics have a common origin, which is the undue objectification of
complex processes, such as those involved in observation. For example, when
experimentalists say they have detected a particle, what they really should
say is that they have done certain things which have had certain observed
consequences. When theorists talk about a cat in a box, they conveniently
ignore the vast number of degrees of freedom needed to define and maintain
the biological state of the cat and talk about it as if it were a particle
with two possible spin states. The fact is, it is possible to have a cat,
parts of which are alive and parts of which are dead in a classical sense.
It is, however, meaningless to talk about a quantum superposition of a
living cat state with a dead cat state without any reference to the
observational meaning of this superposition.

This tendency to use the language of objectification is a very natural one
and frequently leads to meaningless debate about concepts incapable of
verification, such as how many angels can dance on the point of a pin \cite%
{AQUINAS}. Such language is used at all levels in science. Unfortunately, it
is often based on the wrong mode of thinking where quantum mechanics is
involved, because quantum mechanics deals with physics, not metaphysics. By
definition, a metaphysical concept is one which implies a belief about the
world around us, or about the wider universe, which may or may not seem
entirely reasonable but is ultimately not provable or has no possibility of
empirical validation.

We assert that quantum mechanics should be and can be discussed in a
coherent, non-metaphysical way which bypasses the issue of system
objectification. In this paper, we discuss an interpretation and formulation
of quantum mechanics referred to as the QDN (quantized detector network)
approach \cite{1A,1}. This approach is based on treating the quantum
information available to an observer as a primary concept, rather than on
the concepts normally associated with SQM (standard quantum mechanics), such
as Hamiltonians, observables, and states of systems under observation. The
QDN approach may appear extreme from the SQM point of view but the benefits
are immediate; we are no longer constrained to discuss the metaphysical
concept of a system in isolation.

The mathematics of QDN is based on collections of quantum bits forming
networks, the complexity of which is determined by the apparatus involved.
These qubits represent the detectors of quantum information necessarily
associated with any experiment. The result is an approach to quantum
mechanics which not only gives a more faithful description of what actually
happens in the laboratory compared to that given by SQM, but points to new
ways of thinking about and describing physics.

Classical counterfactuality is the notion that properties of objects exist
even if we do not measure those properties. The logic behind this idea is
that if we had decided to measure a property of a system (which in fact we
did not do), then we would have found that property, because it was there
all along. Therefore we do not need to have measured it to justify our
belief in its existence. This principle works wherever classical physics is
valid.

Quantum mechanics, however, teaches us that classical counterfactuality is
wrong in principle. The true laws of physics should be based on the quantum
version of classical counterfactuality, which we shall call quantum
counterfactuality. This is the idea that properties of objects do not exist
independently of the context of observation. This was something Bohr
understood but Einstein never accepted. The EPR thought experiment \cite%
{EPR-1935} was a powerful attempt by Einstein to undermine
quantum mechanics by the subtle use of classical
counterfactuality, and was only countered by Bohr
\cite{BOHR-1935} using the principles of quantum
counterfactuality.

Quantum counterfactuality holds regardless of the temporal context of a
statement. It holds whether we are discussing a situation involving the
past, the present, or the future. It is an incredibly difficult principle to
adhere to in discussions about physics, given our ingrained propensity to
believe in classical counterfactuality. Even our statement of it here uses
conventional language set in classical terms, but this does not detract from
the validity of this principle.

By construction, QDN is based on quantum counterfactuality. QDN should be
able to describe not only everything that the PVM (projection-valued
measure) and POVM (positive operator-valued measure) approaches in SQM can
discuss in more systematic and efficient terms, but also readily deal with
more complex scenarios such as time dependent apparatus, which SQM does not
deal with efficiently. Here, we do not mean simply having a time dependent
potential. We mean for example experiments which change their detecting
apparatus whilst a given run is still in progress.

QDN is an attempt to discuss observation as a dynamical process involving
observers and not as a passive process involving dynamical systems. It is
related to quantum field theory, in that spatial degrees of freedom are
implied in the formalism. In the limit where very many degrees of freedom
are involved in an experiment, a quantum field version of the formalism
should emerge\cite{NOVA-05}. QDN is also related to S-matrix theory {\cite%
{ELOP:1966}}, in that it can be viewed in terms of temporal
sequences of scattering processes. It also bears some
relationship to Schwinger's source theory approach
{\cite{SCHWINGER:1969}}, which is based on a local view of
scattering processes. It is also related to consistent histories theories
\cite{GRIFFITHS-1984}, with the fundamental difference that the histories
involved are those of the apparatus, not of the system under observation and
are always consistent on that account. This aspect of the QDN approach is
discussed in our section of path integrals. Finally, QDN is allied to the
formalism and thinking of quantum information and computation. From the QDN
point of view, any quantum process can be described as a quantum computation.

\section{The Stern-Gerlach experiment}

In order to explain our approach, we start with a brief
discussion of the SG (Stern-Gerlach) experiment \cite{2}. In
this experiment, a beam of charged particles is passed through
an inhomogeneous magnetic field. If the particles are net
carriers of a single electron, then it is observed that a beam
of many such particles splits into two pieces, known as the
\emph{up} beam and the \emph{down} beam respectively. The SQM
description of this phenomenon is due to Pauli \cite{3} and is
based on the fundamental representation of $SU(2)$, the
universal covering group for rotations in
three dimensional physical space. In the two-dimensional Hilbert space $%
\mathcal{H}^{2}$ concerned, an orthonormal basis is chosen, consisting of
one element $|+\mathbf{k}\rangle $ representing the \emph{up} beam particles
and another element $|+\mathbf{k}\rangle $ representing the down beam
particles, where $\mathbf{k}$ is a unit three-vector representing the
spatial direction of the main magnetic field in the apparatus.

The point about using the Hilbert space representation in quantum mechanics
is that vectors can be added. In the absence of any superselection rule, it
is a general rule in quantum mechanics that we can prepare an initial state
of the system (which in this case is usually thought of as a single
electron) represented by the vector
\begin{equation}
|\Psi \rangle =\alpha |+\mathbf{k}\rangle +\beta |-\mathbf{k}\rangle ,
\label{111}
\end{equation}%
where $\alpha $ and $\beta $ are complex and satisfy the normalization
condition $|\alpha |^{2}+|\beta |^{2} = 1$. These squared moduli give the
probabilities of each outcome, according to the Born interpretation of SQM
{\cite{BORN-1926}}.

A number of questions arise at this point. First, there is the issue of
locality. The SG experiment relies on the fact that the up and down beams
intercept the detecting plate at different spatial locations. No account is
taken of this in the superposition (\ref{111}). Therefore, even at this very
basic level, the formalism of SQM exhibits a form of non-locality.

Given this, it is difficult to interpret the superposition (\ref{111}) as
anything more than a metaphor for the experimental procedure. The expansion (%
\ref{111}) has a significance, in that the vectors on the right-hand side
correspond one-to-one with the possible detector sites. This would not
remain true if we now introduced a different orthonormal basis in $\mathcal{H%
}^{2}$, $\left\{ |+\mathbf{a}\rangle ,|-\mathbf{a}\rangle \right\} $, such
that$|+\mathbf{k}\rangle =\frac{1}{\sqrt{2}}\left\{ |+\mathbf{a}\rangle +|-%
\mathbf{a}\rangle \right\} $, $|-\mathbf{k}\rangle =\frac{1}{\sqrt{2}}%
\left\{ |+\mathbf{a}\rangle -|-\mathbf{a}\rangle \right\} $ and rewrote (\ref%
{111}) in the mathematically equivalent form $|\Psi \rangle =\frac{1}{\sqrt{2%
}}(\alpha +\beta )|+\mathbf{a}\rangle +\frac{1}{\sqrt{2}}(\alpha -\beta )|-%
\mathbf{a}\rangle ,$ \emph{unless} we simultaneously rotated the apparatus
so that its main magnetic field direction now pointed along direction $%
\mathbf{a}$, so that $|+\mathbf{a}\rangle $ now represented the new \emph{up}
beam states and $|-\mathbf{a}\rangle $ represented the new \emph{down }%
\noindent beam states.

From such considerations, we feel entitled to infer that the
vector additivity property of Hilbert space has no operational
significance unless reference is made to the apparatus whose
outcomes correspond to the terms in the summation. In other
words, the context of observation (in this case, the direction
of the main magnetic field) is a crucial factor which cannot be
avoided in the quantum formalism. This is in line with the
thinking of Bohr in his famous debate with Einstein on the
interpretation of quantum mechanics \cite{4}. When this point is
neglected, it leads to conceptual issues such as the preferred
basis problem in the relative state/Many Worlds interpretation
of quantum mechanics {\cite{EVERETT-1957}} and theories of
quantum decoherence {\cite{ZUREK-2002}}.

We now come to the main thrust of our approach. At least
\textit{three} questions should be asked in any given run of the
SG experiment. First, we must ask the question "has an electron
been fired from the source or not?". In SQM, it is assumed that
the answer is already \textit{yes}, but in reality, a proper
description of an experiment should involve such a question. We
shall
call this question, and its associated qubit, the \textit{preparation switch}%
. It is a critical feature of the QDN approach, possibly the most important
one, that every qubit in the network has the potential to act as a
preparation switch for some subsequent process.

Next, consider the two potential detector sites, the \emph{up} and \emph{down%
} spots on the SG apparatus detector plate, \emph{just before} a single
electron is detected. At this time, the observer knows that each site
registers nothing. Subsequently, given that an electron has gone through and
registered somewhere in the apparatus (but we are not given where), the
observer would ask the same question at each site, viz., "is there an
electron here or not"? Therefore, contrary to the usual way of describing
the SG experiment as a single question with two possible answers (i.e., spin
\emph{up} or spin \emph{down}), the actual experiment poses \emph{three}
questions, each of which has two possible answers.

The essence of the QDN description of the SG and all other experiments is to
assign a qubit to each place or situation where physicists could in
principle detect new information. These places need not be identifiable with
distinct positions in physical space, but this is often the case. For
instance, the detection of momentum information requires detection apparatus
which is distributed over space in a carefully controlled way, such as in
particle scattering experiments.

The qubits involved in the QDN description are related to each other via
classical information held by the observer, and taken together, this forms
the detector network referred to in our title. The quantum aspect refers to
the time-dependent state of the network, which is represented by a vector,
which we call the \textit{labstate}, in the quantum register formed by the
tensor product of all relevant network qubits. The labstate may be used by
the observer to calculate the Born probability of any possible outcome,
conditional on the answer to the preparation switch question being \textit{%
yes}, prior to any detection. This probability gives the relative
frequencies of the various possible outcomes, if sufficient runs of the
basic experiment are performed.

For the SG experiment, the minimal QDN register is of rank three, i.e.,
involves the tensor product of three qubits, and is therefore a Hilbert
space of dimension eight. This is markedly greater than the two dimensions
used in the SQM Hilbert space description and is the price we have to pay
for having a more complete description of what is going on. To deflect undue
criticism on account of the higher dimensionality, we observe that a full
quantum field theoretic description of even the simplest experiment would
involve an infinite number of degrees of freedom, so our approach is but a
modest departure from the SQM formulation. Moreover, the higher
dimensionality of our Hilbert spaces is useful in modelling more
sophisticated scenarios, such as multiple, staggered runs of an experiment
passing through an apparatus simultaneously. Any apparent redundancy in the
QDN formalism is always capable of physical interpretation.

We include the preparation switch for the following reason. In SQM, state
preparation is regarded as the start of an experiment and state detection is
regarded as the end. In our approach, the two processes are regarded as
synonymous, in that preparation is itself the outcome of some other process,
and detection is but the start of yet another.

In principle we could imagine the physical space between source and
detectors as filled with qubits, but these would be redundant here. In the
QDN approach, we invoke only that number of qubits sufficient to model the
essential physics of a given experiment.

Throughout this paper, we shall suppress the tensor product symbol $\otimes$%
, it being implied. Having introduced a rank-three quantum register $%
\mathcal{R}^{3}\equiv \mathcal{Q}_{0} \mathcal{Q}_{1} \mathcal{Q}_{2}$ to
model the SG experiment, we now discuss a typical run of this experiment.
This involves a time-dependent description of the labstate $|\Psi )$. We
shall denote all labstate vectors using this modification of the Dirac
bra-ket notation, with round angular brackets replacing the traditional
angular brackets, reserving the latter to refer to vectors in the SQM
formalism. Operators acting over labstates will be denoted in blackboard
bold font, such as $\mathbb{U}$. In the following, all references to time
are to laboratory clock time as registered by the observer in the laboratory.

First, imagine the situation after all apparatus has been constructed but
before the actual experiment has started, or between runs. During such a
time, the equipment is lying idle, i.e., unused. It exists, but no electron
is being prepared and no detector is registering any result. Such a state of
the apparatus-laboratory system will be called the \emph{void} state (we
shall not use the term $vacuum$ in this context, as this is generally used
to represent empty space devoid of any matter). We shall represent the void
state by the quantum register vector $|\Psi
_{0})=|0)_{0}|0)_{1}|0)_{2}=|000)=|0.2^{0}+0.2^{1}+0.2^{2})=|0)$.

Suppose now that at some initial time $t_{in}$, a run of the experiment
starts by the process of throwing the preparation switch. At this point the
experimentalists will be confident that the source has prepared an initial
state and that nothing will have been registered yet by either outcome
detector. We represent the labstate at this time by the quantum register
state $|\Psi _{in})=|1)_{0}|0)_{1}|0)_{2}=|100)=|1.2^{0}+0.2^{1}+0.2^{2})
=|1)=\mathbb{A}_{0}^{+}|0)$, where in this particular case $\mathbb{A}%
_{0}^{+}\equiv A_{0}^{+}I_{1}I_{2}$. Here, the operator $A_{0}^{+} =
|1)_{0}(0|$ changes the state of preparation switch from the answer \textit{%
no} to the answer \textit{yes}, whilst $I_{1}$ and $I_{2}$ are the identity
operators acting over $\mathcal{Q}_{1}$ and $\mathcal{Q}_{2}$ respectively.

We can be sure that there must be such an interval of time, because
detectors are generally distinct from sources and therefore can trigger only
at non-zero times \emph{after }state preparation, according to one of the
fundamental principles of special relativity, a principle which we do not
disagree with. All signals registered in our qubits have to be consistent
with the causality properties of relativity.

Towards the end of a given run of the experiment, at a time $t_{out}>t_{in}$%
, the observer may write down the labstate immediately prior to detection,
which is given by
\begin{equation}
|\Psi _{out})=\alpha |010)+\beta |001)=\alpha |2)+\beta |4)=\left( \alpha
\mathbb{A}_{1}^{+}+\beta \mathbb{A}_{2}^{+}\right) |0),  \label{345}
\end{equation}%
where $\mathbb{A}_{1}^{+}\equiv I_{0}A_{1}^{+}I_{2}$ and $\mathbb{A}%
_{2}^{+}\equiv I_{0}I_{1}A_{2}^{+}$.

It is important to understand the origin of the coefficients $\alpha $ and $%
\beta $ in (\ref{345}). In SQM, they are calculated from a knowledge of the
Hamiltonian and other details involved in the experiment. We observe that a
Hamiltonian is not an absolute property of a system under observation, even
though it is customary to talk about the "Hamiltonian of a system". It
changes according to the context of the experiment. For instance, changing
the magnetic field in the SG experiment changes the Hamiltonian. Therefore,
even in SQM, what is usually referred to as the Hamiltonian of the system is
in practical terms determined by the apparatus. It is only by virtue of
classical knowledge previously acquired by an observer about their apparatus
that permits a discussion if terms of a Hamiltonian for a system.

Given a Hamiltonian as a depository of acquired knowledge about an
experiment, SQM then processes this knowledge via standard rules. In QDN,
this knowledge will be represented and processed in a different though
equivalent format, one designed to emphasize the apparatus rather than the
system. In both SQM and QDN, the ambition is essentially the same however,
i.e., to write down coefficients which carry information about the outcome
probabilities of an experiment.

A practical difficulty with QDN emerges here. Currently, our experience of
QDN is relatively limited. We are working our way towards a comprehensive
dynamical theory, which we envisage will give us consistent methods of
dynamical calculations entirely within a QDN framework. At this time,
however, we have not yet developed the formalism sufficiently to allow us a
method of calculating the coefficients $\alpha $ and $\beta $ without
resorting at least somewhere to a knowledge of SQM. It would not be correct
to criticize the QDN approach on that account, as the importance of the
conceptual issues transcends the significance of the practical difficulties
we face at this time.

Given $|\Psi _{out}),$ the Born probability rule adapted to the
quantum register can be applied to give the outcome
probabilities
\begin{eqnarray} P\left( \textrm{up}|\Psi
_{in}\right) &\equiv &|(2|\Psi _{out})|^{2}=|\alpha |^{2},\\
P\left( \textrm{down}|\Psi _{in}\right) &\equiv& |(4|\Psi
_{out})|^{2}=|\beta |^{2},  \nonumber \\ P\left( \textrm{any
other state}|\Psi _{in}\right) &\equiv &|\left( a|\Psi
_{out}\right) |^{2}=0,\;\;\;a=0,1,3,5,6,7,
\end{eqnarray}%
consistent with known physics. Of course, during any single run involving a
single electron, only one detector gets triggered, so these probabilities
have to be related to the frequencies of outcome built up over many runs of
the basic experiment.

Normally, after each run is over and before the next one starts, the
labstate reverts to the void state $|0)$. The specific mechanism for this is
currently beyond known physics, as is the transition from the void state to
the initial state at the start of a run. These end-point transitions involve
extremely complex processes, such as those associated with the observers
themselves and the factors involved in their decision making actions.
Virtually nothing is known about this side of physics and it is ignored in
SCM on that account, but it is remarkable that Feynman recognized this as an
issue worthy of comment. In an influential article on the simulation of
physics with computers, he wrote \cite{FEYNMAN-1982} \textit{%
\textquotedblleft ...we have an illusion that we can do any experiment that
we want. We all, however, come from the same universe, have evolved with it,
and don't really have any \textquotedblleft real\textquotedblright\ freedom.
For we obey certain laws and have come from a certain past.\textquotedblright%
}

We envisage that any experiment currently devised could be embedded as part
of a bigger network, in such a way that what are perceived as state
preparation and outcome detection processes are simply parts of the greater
network in action. In essence, a laboratory experiment is no more or less
than a particular local manifestation of the universe running as a vast
quantized detector network. Feynman's concern about why a given experiment
was done in the first place would then find an answer somewhere within the
principles governing that greater network. This, however, would require a
much greater understanding of large scale quantized networks than we have at
present.

\section{Formal developments}

We may consider changes in the labstate during a run of the SG experiment to
be described by unitary evolution over the quantum register $\mathcal{R}^{3}$%
, viz, $|\Psi _{in})\rightarrow |\Psi _{out})\equiv \mathbb{U}\left(
t_{out},t_{in}\right) |\Psi _{in})$, where $\mathbb{U}\left(
t_{out},t_{in}\right) $ is unitary so as to preserve total probability.
Exactly what such an operator is or should be will not always be known,
because physics experiments will not in general deal with absolutely every
possible state in a quantum register. The basic SG experiment, for example,
requires us only to consider four of the eight computational basis elements,
viz, $|0),|1),|2)$ and $|4)$. It says nothing about the states involving the
basis elements $|3),|5),|6)$ and $|7)$. Even for such basic experiments as
the SG experiment, there is a degree of overkill in the QDN description.
This should not be regarded as a flaw however; an analogous situation occurs
in all classical and quantum theories. The point is, we only need to know
the action of $\mathbb{U}\left( t_{out},t_{in}\right) $ on certain states in
order to make empirical predictions.

In this context, we can be sure of one or two things with a degree of
confidence. We can be confident that in between state preparation and
outcome detection, time evolution conserves probability. Also, if the
apparatus is in a void state, then we do not expect that to change, unless
we initiate a new run (which will not conserve probability anyway).
Therefore, we may assume $\mathbb{U}\left( t_{out},t_{in}\right) |0)=|0)$.
Then for the SG experiment, for example, we can represent the dynamics in
terms of how the transition operators change, viz,
\begin{equation}
\mathbb{A}_{0}^{+}\rightarrow \mathbb{U}\left( t_{out},t_{in}\right) \mathbb{%
A}_{0}^{+}\mathbb{U}^{+}\left( t_{out},t_{in}\right) =\alpha \mathbb{A}%
_{1}^{+}+\beta \mathbb{A}_{2}^{+}.
\end{equation}%
More generally, we shall \textquotedblleft modularize\textquotedblright\ our
spatio-temporal description, meaning that we shall discuss how individual
transition operators change at various times in their own ways. Often we
shall leave out specific reference to the $\mathbb{U}$ operators, writing
for example $\mathbb{A}_{0}^{+}\rightarrow \alpha \mathbb{A}_{1}^{+}+\beta
\mathbb{A}_{2}^{+}$ to describe a particular change in the operator $\mathbb{%
A}_{0}^{+}$ at a particular place and time during a given run.

Because real physics experiments are irreversible, we need to be cautious
about what operators such as $\mathbb{U}\left( t_{out},t_{in}\right) $
really mean. They will have the semi-group property $\mathbb{U}\left(
t_{2},t_{1}\right) \mathbb{U(}t_{1},t_{0}\mathbb{)=U(}t_{2},t_{0}\mathbb{)}$%
, $t_{2}\geqslant t_{1}\geqslant t_{0}$ and satisfy the rule $\mathbb{U}%
\left( t_{1},t_{0}\right) \mathbb{U}^{+}\left( t_{1},t_{0}\right) =\mathbb{I}%
_{\mathcal{R}}$, where $\mathbb{I}_{\mathcal{R}}$ is the register identity
operator, but we need have no clear physical interpretation of what the
operator $\mathbb{U}\left( t_{0},t_{1}\right) $ means, for $t_{0}<t_{1}$.
Time reversal in QDN is regarded as meaningful only as a comparison between
different but related networks operating forwards in time, and not in terms
of any metaphysical running of a network backwards in time. The apparent
time reversal properties of SQM arise only when fundamental issues to do
with apparatus are ignored.

\section{von Neumann tests}

The SG experiment is the most elementary and useful example of
the sort of quantum experiment discussed by von Neumann
\cite{VON-NEUMANN:1955}, where an ensemble of identically
prepared initial states is passed one at a time through some
test apparatus $A$ and a range of possible outcomes detected.
The description of an idealized version of such an experiment
leads to the so-called projection-valued measure (PVM)
description of quantum experiments. This is known to have its
limitations, but remains an important concept.

The general PVM scenario goes as follows. For each run of an ensemble of
runs, the initial state $|\Psi _{in}\rangle $, which will be assumed to be
pure, is prepared by some apparatus $\Sigma _{0}$ at time $t_{in}$.
Subsequently, the prepared state is passed through test apparatus $A$, and
one out of $d$ possible outcomes detected at time $t_{out}$. In von
Neumann's approach, $|\Psi _{in}\rangle $ is assumed to be a normalized
element of some $d-$dimensional Hilbert space $\mathcal{H}$ and the test $A$
is represented by some non-degenerate Hermitian operator $\hat{A}$ acting
over $\mathcal{H}$. Because of non-degeneracy, the eigenstates $%
|a_{1}\rangle $, $|a_{2}\rangle $,$\ldots ,|a_{d}\rangle $ can be normalized
and form an orthonormal basis for $\mathcal{H}$, known as the preferred
basis.

Because of completeness, we may write
\begin{equation}
|\Psi _{in}\rangle \rightarrow |\Psi _{out}\rangle =\hat{U}\left(
t_{out},t_{in}\right) |\Psi _{in}\rangle =\sum_{i=1}^{d}\Psi
^{i}|a_{i}\rangle ,
\end{equation}%
where
\begin{equation}
\Psi ^{i}=\langle a_{i}|\Psi _{out}\rangle =\langle a_{i}|\hat{U}\left(
t_{out},t_{in}\right) |\Psi _{in}\rangle .  \label{DDD}
\end{equation}%
The Born probability interpretation then predicts the conditional outcome
probabilities to be given by $P\left( a_{i}|\Psi _{in}\right) =|\langle
a_{i}|\Psi _{out}\rangle |^{2}=|\Psi ^{i}|^{2}.$

The QDN description of the PVM scenario follows the pattern outlined for the
SG experiment above. We associate one qubit with every part of the apparatus
wherever a state could be detected and new information acquired. This means
one qubit for the preparation switch and one for each of the $d$ possible
outcomes. Therefore, we need a rank-$\left(1+d\right) $ quantum register for
such a test.

The QDN dynamics is given by the rule $\mathbb{A}_{0}^{+}\rightarrow
\sum_{i=1}^{d}\Psi ^{i}\mathbb{A}_{i}^{+}$, where the $\Psi ^{i}$ are given
by the conventional quantum calculation (\ref{DDD}), so we find $|\Psi
_{in})\rightarrow |\Psi _{out})=\sum_{i=1}^{d}\Psi ^{i}\mathbb{A}%
_{i}^{+}|0)=\sum_{i=1}^{d}\Psi ^{i}|2^{i})$. The conditional probabilities
for the $d$ possible outcomes of the experiment are then given by the
quantum register Born rule $P\left( a_{i}|\Psi _{in}\right) \equiv |\left(
2^{i}|\Psi _{out}\right) |^{2}=|\Psi ^{i}|^{2}$, in agreement with the PVM
formalism.

If all experiments were of this form, there would be little
advantage in the QDN description. This comes into its own when
more than one von Neumann test are coupled together, either in
series, parallel, or a combination of both. This is a situation
which occurs frequently in quantum optics experiments and which
we have discussed in some detail \cite{1}. Another scenario is
slit experiments, which we discuss next.

\section{Slit experiments}

A slit experiment is one where a particle source channels a beam
of particles onto two or more openings in an otherwise opaque
barrier. Classically, these openings are associated with
mutually exclusive pathways for single particles. On the other
side of the barrier is a detecting screen, which registers, at
local sites, places where individual particles have landed after
passage through the slit-barrier. Examples of such experiments
are photon diffraction experiments conventionally called Young's
double slit experiment and electron diffraction experiments of
both the analogue (Davisson and Germer) type
\cite{DAVISSON+GERMER-1927} and modern digitized versions of
them.

We shall describe a general version of this scenario in the QDN
representation as follows. The source is represented as before by single
qubit $\mathcal{Q}_{0}$, the preparation switch. Each slit in the barrier is
in principle a place where a particle detector could be placed, and
therefore, we introduce a qubit $\mathcal{Q}_{B,a}$ for each slit. Here the
subscript $B$ refers to "barrier" and the index $a$ is an integer running
from $-\infty $ to $+\infty $. On the detecting screen, there will be a
countable number of sites on which the particles can land, and so we
introduce a qubit $\mathcal{Q}_{D,j}$ for each such site. Here the subscript
$D$ stands for "detector". In real experiments involving say photographic
film, there will always be a finite number of sites, not a continuum of
sites as represented in SQM. In principle, because we could make the
detecting screen as large as we want, we shall allow the index $j$ on the
detecting qubits to run from $-\infty $ to $+\infty $, as for the slit
qubits.

A typical run of the experiment starts, as before, with the apparatus in the
void state $|0)\equiv |0)_{0}\Pi _{i=-\infty }^{\infty }|0)_{B,i}\Pi
_{j=-\infty }^{\infty }|0)_{D,j}$. Then at time $t=0$, a switch is pulled
and a particle is emitted from the source. The labstate in now given by $%
|\Psi _{in})$ $=\mathbb{A}_{0}^{+}|0)$.

In the conventional scenario, it is imagined that a Schr\"{o}dinger wave
impinges on the barrier and is split into as many parts as there are slits
in it. Each part of the wave then passes through a single slit in the
barrier and interference with the other parts takes places on the other side
of the barrier to the source. In the QDN representation, the splitting of
the wave by the slits in the barrier is represented by the unitary evolution
process $\mathbb{A}_{0}^{+}\rightarrow \mathbb{U}_{0}\mathbb{A}_{0}^{+}%
\mathbb{U}_{0}^{+}\equiv \sum_{a=-\infty }^{\infty }\Psi ^{a}\mathbb{A}%
_{B,a}^{+}$, where the coefficients $\Psi _{a}$ depend on the details of the
experimental arrangement, but must satisfy the rule%
\begin{equation}
\sum_{a=-\infty }^{\infty }|\Psi ^{a}|^{2}=1  \label{222}
\end{equation}%
in order to conserve probability. This part of the experiment will be
referred to as splitting.

The evolution subsequent to splitting, between the barrier and the detector,
is where quantum interference takes place. This means, in concrete terms,
that the separate beams do not interact classically with each other, apart
from superposing, thereby creating a final, superposed wave. In the QDN
representation, this part of the experiment is described using two rules.
First, each slit behaves as if it were now a preparation switch, i.e., a
source of a new PVM experiment, with a collection of detectors at the
detector screen. Hence we write $\mathbb{A}_{B,a}^{+}\rightarrow \mathbb{U}%
_{1}\mathbb{A}_{B,a}^{+}\mathbb{U}_{1}^{+}\equiv \sum_{j=-\infty }^{\infty
}U_{aj}^{(1)}\mathbb{A}_{D,j}^{+}$, where again, unitarity requires us to
impose the conditions%
\begin{equation}
\sum_{j=-\infty }^{\infty }U_{aj}^{(1)\ast }U_{bj}^{(1)}=\delta _{ab}.
\label{333}
\end{equation}%
To prove this, we note that the "slit" states $\mathbb{A}_{B,a}^{+}|0)$ are
mutually orthogonal states in the QDN register, and we simply apply unitary
evolution within the left hand side of the inner product $(0|\mathbb{A}_{B,a}%
\mathbb{A}_{B,b}^{+}|0)=\delta _{ab}$.

The second rule involves superposition. Just prior to detection at the
detector screen, the labstate is given by%
\begin{equation}
|\Psi _{out})\equiv \mathbb{U}_{1}\mathbb{U}_{0}\mathbb{A}_{0}^{+}\mathbb{U}%
_{0}^{+}\mathbb{U}_{1}^{+}|0)=\sum_{a=-\infty }^{\infty }\sum_{j=-\infty
}^{\infty }\Psi ^{a}U_{aj}^{(1)}\mathbb{A}_{D,j}^{+}|0),
\end{equation}%
from which we read off the transition amplitude $A_{j}$ of a particle
landing on detector $j$ to be given by $A_{j}=\sum_{a=-\infty }^{\infty
}\Psi ^{a}U_{aj}^{(1)}$. Hence the probability $P_{j}$ that the detector
associated with qubit $\mathcal{Q}_{D,j}$ fires, conditional on the
experiment as set up, is given by $P_{j}=|A_{j}|^{2}$. It is easy to verify,
using the summation rules (\ref{222}) and (\ref{333}) that these
probabilities sum to unity.

In principle, there is no need to visualize the barrier or detecting screens
in terms of standard geometrical objects, such as planes. The barrier and
detector sites could be arranged in any sort of spatial configuration
whatsoever. It is the coefficients $\left\{ \Psi ^{a}\right\} $ and the $%
\left\{ U_{aj}^{(1)}\right\} $ which determine whether any geometrical
interpretation is meaningful. These coefficients are, in line with the
interpretation we advocate throughout this paper, manifestations of the
physics of the apparatus, not of any imagined system passing through the
apparatus.

However, there will be situations where classical information about the
distribution of sites within the apparatus permits the use of symmetry
arguments. For example, suppose the experiment does indeed involve screens,
with the slits equally spaced along a given direction in the barrier plane,
and with the detector sites equally spaced along the same direction in the
detector plane. If the slits are of equal width and the detectors identical
in construction, then spatial homogeneity may allow us to write $%
U_{aj}^{(1)}\equiv V_{a-j}$, where the complex coefficients $\left\{
V_{k}\right\} $ satisfy the rule $\sum_{k=-\infty }^{\infty }V_{k}^{\ast
}V_{k+d}=\delta _{0,d}$.

A double slit experiment is a particular version of the above setup such
that all but two of the slits are blocked, with the unblocked slits placed
in a symmetrical way relative to the source. Suppose the slits labelled by
positive integer $s$ and its counterpart $-s$ are unblocked. Then up to the
point of splitting, QDN evolution is given by $\mathbb{A}_{0}^{+}\rightarrow
\Psi ^{s}\mathbb{A}_{B,s}^{+}+\Psi ^{-s}\mathbb{A}_{B,-s}^{+},$ where in the
symmetrical case, $|\Psi ^{s}|^{2}=|\Psi ^{-s}|^{2}=1/2$. Subsequently and
just prior to detection, evolution gives%
\begin{equation}
\mathbb{A}_{0}^{+}\rightarrow \sum_{j=-\infty }^{\infty }\left\{ \Psi
^{s}V_{s-j}+\Psi ^{-s}V_{-s-j}\right\} \mathbb{A}_{D,j},
\end{equation}%
so that
\begin{eqnarray}
P_{j} &=&|\Psi ^{s}|^{2}|V_{s-j}|^{2}+|\Psi ^{-s}|^{2}|V_{-s-j}|^{2}
\nonumber \\
&&+\Psi ^{-s}\Psi ^{s\ast }V_{-s-j}V_{s-j}^{\ast }+\Psi ^{-s\ast }\Psi
^{s}V_{-s-j}^{\ast }V_{s-j}.
\end{eqnarray}%
These probabilities sum to unity as required. We see here the appearance of
interference terms regardless of the details of the dynamics.

\section{Path integrals}

In SQM, the Feynman path integral approach
{\cite{FEYNMAN+HIBBS:1965}} utilizes the full power of quantum
counterfactuality to rewrite the amplitude for a given initial
state of a system to evolve to some final state in terms of all
possible paths between the two states. We show in this section
how the QDN approach can be recast in path integral terms.
Feynman's concept of all possible paths contributing to the
amplitude is essentially identical to that of a quantized
detector network, with the important difference that the
emphasis in Feynman's approach is on states of the system and
not on those of the apparatus.

Up to now, the QDN approach has been to suppose that detector qubits are
fixed throughout an experiment. This need not be the case in reality. We
could imagine an experiment lasting over such a duration that whilst it was
still in progress, parts of the apparatus were being constructed or removed.
If we went so far as to imagine that the universe itself is some sort of
vast quantum process, then this phenomenon of time-dependent apparatus
emerges logically. It leads inevitably to the concept of a quantum cosmology
where the universe is regarded as a self-interacting, autonomous quantum
dynamical system with no external observers. An example of what we mean
comes from astrophysics. Astronomers occasionally receive light from a
supernova, which shows in the most graphical terms that the source of the
light no longer exists in the same form, that of an unexploded star, that it
had just prior to the emission of that light.

We do not have the space to discuss this side of the QDN
formalism further here, save to make three comments. First, the
construction of a consistent, mathematically well-defined
quantum cosmology was one of the factors which motivated our
ideas \cite{NOVA-05}. Second, the concept of apparatus changing
dynamically is relevant to what happens normally in the world
about us. Because of the extreme complexity involved in such
changes, however, physicists generally go to great lengths to
construct their experiments so as to eliminate as much as
possible the effects of apparatus change during the course of
their experiments. The result is that in SQM, apparatus is
assumed to be fixed, but we emphasize that this is no more than
a useful assumption. The QDN approach allows us to go beyond
this scenario. Third, we believe that the quantization of
general relativity should be tackled through a time-dependent
QDN approach or its equivalent, because space, time and metric
are inherently aspects of the process of observation and not
intrinsic properties of systems under observation.

We shall calculate the amplitude for a QDN to go from an initial labstate at
time $0$ to some final labstate at time $N$, where $N$ is some positive
integer. In our approach, time is necessarily discrete, because the process
of information extraction is never instantaneous (the change from the
concept of continuous time to discrete time is a logical one once we move
away from systems and discuss matters in observational terms). Therefore,
the labstate associated with the apparatus will be assumed to evolve in a
finite sequence of steps, called stages, each labelled by an integer $n$
running from zero to $N$. The physical time between successive stages need
not be uniform, however, or small, such as on Planck scales. We note in
passing that the Feynman path integral is usually formulated via a
discretization of time, and that the continuum limit is generally
ill-defined, if not formally non-existent. We have no such problem in the
QDN approach, because the continuous time limit is not one that we believe
is strictly in accordance with the observational facts of quantum
experiments. We believe that the continuous time concept is meaningful only
in so far as there are virtually limitless degrees of freedom associated
with the observer. In the scenario mentioned above, concerning Feynman's
statement about the freedom of choice in an experiment, we imagine any
experiment being part of a greater network described by vastly many qubits,
and then a description of that greater network might well be more efficient
in terms of continuous variables. There is an analogy here with the
relationship of classical thermodynamics to statistical mechanics.

Taking into account the idea that the qubits associated with the apparatus
at time $n$ may be different to those at time $n+1$, we shall label the
operators and labstate associated with time $n$ with the index $n$. We shall
suppose that there are $r_{n}$ qubits involved at time $n$.

At time zero, after the preparation switch has been thrown for qubit $%
\mathcal{Q}_{0,i_{0}}$, $1\leqslant i_{0}\leqslant r_{0}$, the labstate is
given by $|\Psi _{0},i_{0})\equiv \mathbb{A}_{0,i_{0}}^{+}|0)$, where $%
1\leqslant i_{0}\leqslant r_{0}$. At the next stage, the labstate is given
by the transition rule $|\Psi _{0},i_{0})\rightarrow \mathbb{U}_{1}|\Psi
_{0},i_{0})=\sum_{i_{1}=1}^{r_{1}}U_{i_{0}i_{1}}^{(1)}\mathbb{A}%
_{1,i_{1}}^{+}|0)$, where the complex coefficients $\left\{
U_{i_{0}i_{1}}^{(1)}\right\} $ satisfy the "unitarity" rule
\begin{equation}
\sum_{i_{1}=0}^{r_{1}}U_{i_{0}i_{1}}^{\left( 1\right)
}U_{j_{0}i_{1}}^{\left( 1\right) \ast }=\delta _{i_{0}j_{0}},\ \ \
1\leqslant i_{0},j_{0}\leqslant r_{0}.
\end{equation}%
There is no reason to expect $r_{1}$ to be necessarily equal to $r_{0}$, so
the matrix $U_{ij}^{\left( 1\right) }$ is not necessarily square, and
therefore, not necessarily a unitary matrix in the conventional sense.

At time $t=1$, if no measurement is taken, each of the qubits involved with
the labstate at that time becomes a preparation switch for the next step of
the evolution. Therefore, we simply iterate the process in the same way,
giving%
\begin{equation}
|\Psi _{0},i_{0})\rightarrow \mathbb{U}_{2}\mathbb{U}_{1}|\Psi
_{0},i_{0})=\sum_{i_{1}=1}^{r_{1}}%
\sum_{i_{2}=1}^{r_{2}}U_{i_{0}i_{1}}^{(1)}U_{i_{1}i_{2}}^{(2)}\mathbb{A}%
_{2,i_{2}}^{+}|0),
\end{equation}%
where the coefficients $\left\{ U_{i_{1}i_{2}}^{(2)}\right\} $ satisfy a
corresponding unitarity rule.

Taking this process all the way to time $N$ gives us%
\begin{equation}
|\Psi _{0},i_{0})\rightarrow \mathbb{U}_{N}\mathbb{U}_{N-1}\ldots \mathbb{U}%
_{1}|\Psi _{0},i_{0})=\sum_{i_{N}=1}^{r_{N}}\ldots
\sum_{i_{2}=1}^{r_{2}}U_{i_{0}i_{1}}^{(1)}\ldots U_{i_{N-1}i_{N}}^{(N)}%
\mathbb{A}_{N,i_{N}}^{+}|0).
\end{equation}%
From this, we read off the amplitude $\mathcal{A}\left( i_{0},i_{N}\right) $
for the labstate to go from $\mathbb{A}_{0,i_{0}}|0)$ to $\mathbb{A}%
_{N,i_{N}}|0)$ to be given by%
\begin{equation}
\mathcal{A}\left( i_{0},i_{N}\right) =\sum_{i_{N-1}=1}^{r_{N-1}}\ldots
\sum_{i_{1}=1}^{r_{2}}U_{i_{0}i_{1}}^{(1)}\ldots U_{i_{N-1}i_{N}}^{(N)}.
\end{equation}%
This is our path integral formulation of QDN dynamics.

The above formulation is not the most general conceivable. It is
based on the detector network changing in a predetermined,
classical manner independent of the labstate. A more subtle
concept would be to link the changes in the network qubits with
changes in the labstate \cite{JAG-2005}. This gives an
altogether more dynamical view of the universe. It is our belief
that something like this is the proper way to discuss quantum
gravitation.

\section{The POVM formalism}

The PVM formulation of quantum physics eventually became superseded by the
more general POVM (positive operator-valued measure) approach \cite%
{BRANDT-1999}. In this latter approach, quantum experiments can have more or
less outcomes than the dimension of the Hilbert space involved in the
description of the prepared state. For example, suppose we have an
experiment with $k$ possible outcomes, with $k$ not necessarily equal to $d$%
, the dimension of the Hilbert space $\mathcal{H}$ used to model the states
of the system. For each outcome $|\phi ^{i}\rangle $, $i=1,2,\ldots ,k$,
there is an associated positive operator $\hat{E}_{i}$, such that
\begin{equation}
\sum_{i=1}^{k}\hat{E}_{i}=\hat{I}_{\mathcal{H}},  \label{BBB}
\end{equation}%
where $\hat{I}_{\mathcal{H}}$ is the identity operator over the Hilbert
space. Given a normalized initial state $|\Psi \rangle \in \mathcal{H}$,
then the conditional probability $P\left( \phi ^{i}|\Psi \right) $ of
outcome $|\phi ^{i}\rangle $ is given by $P\left( \phi ^{i}|\Psi \right)
=\langle \Psi |\hat{E}^{i}|\Psi \rangle $, with condition (\ref{BBB})
ensuring probabilities sum to unity.

The above discussion involves pure states. The POVM approach is not
restricted to these and can be extended to cover mixed states, which
requires a density matrix approach involving the taking of traces. We shall
not discuss mixed states in this paper, as the generalization of QDN to
cover such cases is not anticipated to be particularly difficult and is left
as an exercise for the reader.

The disadvantage of the POVM approach is that it masks the spatio-temporal
structure of the measurements involved and suggests that the simple in-out
architecture of a single von Neumann test is all that is going on. In
reality, complex experiments involve sequences of processes rather like
computations in a computer, which is why quantum computation is one possible
way to approach physics \cite{FEYNMAN-1982}.

The QDN approach readily deals with situations discussed by the
POVM formalism in SQM. We have shown \cite{1} how readily a
quantum optics
experiment discussed recently by Brandt using a POVM description \cite%
{BRANDT-1999} can be discussed using QDN, with the advantage that it is
conceptually more understandable than the POVM approach. In particular, all
of the detector qubits are treated in the same way, whereas the status of
the detectors in the POVM approach is considered inequivalent. This is
because two of the three detectors involved project onto non-orthogonal
states in the SQM Hilbert space involved. The QDN description of the
experiment makes it clear that the original formulation of the experiment in
terms of non-orthogonal basis vectors is not necessary, and is induced by
the choice of apparatus.

\section{Higher rank states}

A feature of the formalism developed thus far is that, apart from the void
state, all physical states discussed have been of the form $|2^{k})\equiv
\mathbb{A}_{k}^{+}|0)$ or linear combinations of them. Such states will be
called \emph{rank-one states}. We generalize the concept of state rank as
follows.

Given a rank-$r$ quantum register $\mathcal{R}^{r}\equiv \mathcal{Q}_{0}%
\mathcal{Q}_{1}\mathcal{Q}_{2}\ldots \mathcal{Q}_{r-1}$ and its associated
computational basis $B^{r}\equiv \left\{ |a):0\leqslant a<2^{r}\right\} ,$
we define the \emph{rank-}$p$\emph{\ subsets} $B_{p}^{r}$ of $B^{r}$ as
follows: $B_{0}^{r}\equiv \left\{ |0)\right\} $, $B_{1}^{r}\equiv \left\{
\mathbb{A}_{a}^{+}|0):0\leqslant a<2^{r}\right\} $, $B_{2}^{r}\equiv \{%
\mathbb{A}_{a}^{+}\mathbb{A}_{b}^{+}|0):0\leqslant a<b<2^{r}\},\ldots
,B_{r}^{r}\equiv \left\{ \mathbb{A}_{0}^{+}\mathbb{A}_{1}^{+}\ldots \mathbb{A%
}_{r-1}^{+}|0)\right\} $. There are $r+1$ such subsets.

The cardinality $\#B_{p}^{r}$ of $B_{p}^{r}$ is given by
$\#B_{p}^{r}= \frac{r!}{\left( r-p\right)
!p!}$. The rank-$p$ subsets are
disjoint and exhaustive, i.e., $B_{p}^{r}\cap B_{q}^{r}=\emptyset $ for $%
p\neq q,$ and $B_{0}^{r}\cup B_{1}^{r}\cup \ldots \cup B_{r}^{r}=B^{r}$. The
elements of $B_{p}^{r}$ are linearly independent vectors in $\mathcal{R}^{r}$
and therefore form an orthonormal basis for a vector subspace of dimension
$\#B_{p}^{r}$, which we shall denote by $\mathcal{R}_{p}^{r}$.
Any vector in $\mathcal{R}_{p}^{r}$ will be called a rank-$p$
state. Each rank-$p$ subspace $\mathcal{R}_{p}^{r}$ is a
bona-fide Hilbert space, but apart from the trivial subspaces
$\mathcal{R}_{0}^{r}$ and $\mathcal{R}_{r}^{r}$, is \textit{not}
a quantum register.

Different level subspaces have only the zero vector $\mathbf{0}_{\mathcal{R}}
$ of $\mathcal{R}^{r}$ in common, i.e., $\mathcal{R}_{p}^{r}\cap \mathcal{R}%
_{q}^{r}=\{\mathbf{0}_{\mathcal{R}}\mathbf{\}}$, $p\neq q$, and so we deduce
$\mathcal{R}_{0}^{r}\oplus \mathcal{R}_{1}^{r}\oplus \ldots \oplus \mathcal{R%
}_{r}^{r}=\mathcal{R}^{r},$ where $\oplus $ denotes the direct sum of the
subspaces concerned.

\section{Interpretation of higher rank labstates}

There are important situations in QDN physics where labstates of rank higher
than one are encountered naturally. We discuss some of these next.

\subsection{Independent experiments}

Suppose two SG experiments, $A$ and $B$, are performed separately and
completely independently of each other in different parts of the universe.
In such a case, we can describe the two experiments by a single rank-$6$ QDN
involving rank-2 states as follows. First, we assign qubits $\mathcal{Q}_{0}$%
, $\mathcal{Q}_{1}$ $\mathcal{Q}_{2}$ to experiment $A$ and qubits $\mathcal{%
Q}_{3}$, $\mathcal{Q}_{4}$ and $\mathcal{Q}_{5}$ to experiment $B$. Now
suppose experimentalists at each of the two laboratories have agreed to run
separate runs of their experiments simultaneously. Then the initial labstate
of the simultaneous experiments over the rank-6 network is given by
\begin{equation}
|\Psi _{in})=\mathbb{A}_{0}^{+}\mathbb{A}%
_{3}^{+}|0)=|100100)=|2^{0}+2^{3})=|9).
\end{equation}%
If now each experiment is truly independent, then the dynamics followed by
the qubits associated with each experiment are independent, and so we can
write $\mathbb{A}_{0}^{+}\rightarrow \alpha \mathbb{A}_{1}^{+}+\beta \mathbb{%
A}_{2}^{+}$ and $\mathbb{A}_{3}^{+}\rightarrow \gamma \mathbb{A}%
_{4}^{+}+\delta \mathbb{A}_{5}^{+}$, where $|\alpha |^{2}+|\beta
|^{2}=|\gamma |^{2}+|\delta |^{2}=1$. Hence
\begin{equation}
|\Psi _{in})\rightarrow |\Psi _{out})=\left( \alpha \mathbb{A}_{1}^{+}+\beta
\mathbb{A}_{2}^{+}\right) \left( \gamma \mathbb{A}_{4}^{+}+\delta \mathbb{A}%
_{5}^{+}\right) |0)=|\psi )_{A}\otimes |\phi )_{B},
\end{equation}%
where $|\psi )_{A}\equiv \alpha |0)_{0}|1)_{1}|0)_{2}+\beta
|0)_{0}|0)_{1}|1)_{2}$ and $|\phi )_{B}\equiv \gamma
|0)_{3}|1)_{4}|0)_{5}+\delta |0)_{3}|0)_{4}|1)_{5}.$ In other words,
independent experiments are modelled in QDN by separable states of rank
higher than unity over a single (universal) network.

In such scenarios, sets of coefficients such as $\left\{ \alpha ,\beta
\right\} $ and $\left\{ \gamma ,\delta \right\} $ associated with separate
experiments will be functionally independent if the associated experiments
are truly independent. This means in practical terms that if the
experimentalists associated with $A$ were to change the orientation of the
main magnetic field in their apparatus, thereby changing $\alpha $ and $%
\beta $, there would be no change in $|\phi )_{B}$. The same would hold for $%
B$. If the experiments were too close, however, we might imagine that the
magnetic field of one would overlap that of the other, so that changes in
one field would be detected in the other apparatus. The two sets of
coefficients would then no longer be independent.

The possibility of discussing independent quantum processes within a
universal QDN by a separable labstate is complemented by the possibility of
having entangled labstates. For example, we could construct an experiment
where the output channels of $A$ and $B$ in the above were separable but
were subsequently fed into a third apparatus $C$ which had only entangled
outcomes. Particle scattering experiments are of this type.

\subsection{Change of rank experiments: (EPR)}

Experiments of the type discussed by Einstein, Podolsky and Rosen \cite%
{EPR-1935} cause conceptual problems because they confront the classical
system concept with quantum non-locality. The QDN approach avoids this
problem by simply not referring to the system concept at all.

Suppose we prepared a spin-zero bound state of an electron and a positron,
given in SQM by $|\Psi \rangle =2^{-1/2}%
\left\{ |+\mathbf{%
k\rangle }_{e}|-\mathbf{k\rangle }_{p}-|-\mathbf{k\rangle }_{e}|+\mathbf{%
k\rangle }_{p}\right\} $, where the subscripts $e$ and $p$ refer to
electron and positron respectively. Alice and Bob are two well-separated
observers, each with their own particle species filters and SG equipment.
Alice can detect and test for electron spin only, whereas Bob can detect and
test for positron spin only. Alice sets her quantization axis along $\mathbf{%
k}=(0,0,1)$, whereas Bob sets his along direction $\mathbf{a}=(\sin \theta
\cos \phi ,\sin \theta \sin \phi ,\cos \theta )$. It is known that whenever
Alice finds that an electron has passed through her apparatus with spin $|+%
\mathbf{k\rangle }$, Bob will have found his positron had passed through his
apparatus via either of the $|+\mathbf{a\rangle }$ or $|-\mathbf{a\rangle }$
channels in a random way, with frequency given correctly by quantum
mechanics.

The QDN description requires a minimum of five qubits; one for the
preparation switch, two for Alice's SG apparatus outcomes and two for those
of Bob's apparatus. To determine the QDN dynamical rules, we look at the SQM
calculation. We recall that the standard rules permit us to rewrite the
prepared state in the form
\begin{equation}
|\Psi \rangle =\frac{1}{\sqrt{2}}\left\{
\begin{array}{c}
\sin (\frac{_{1}}{^{2}}\theta )e^{-i\phi }|+\mathbf{k\rangle }_{e}|+\mathbf{%
a\rangle }_{p}+\cos (\frac{_{1}}{^{2}}\theta )e^{-i\phi }|+\mathbf{k\rangle }%
_{e}|-\mathbf{a\rangle }_{p} \\
-\cos (\frac{_{1}}{^{2}}\theta )|-\mathbf{k\rangle }_{e}|+\mathbf{a\rangle }%
_{p}+\sin (\frac{_{1}}{^{2}}\theta )|-\mathbf{k\rangle }_{e}|-\mathbf{%
a\rangle }_{p}%
\end{array}%
\right\} .
\end{equation}%
This is the form which relates best to the experiment as actually done. We
use this form to determine the transition rules for the labstate. We require
the QDN dynamics to satisfy the rule%
\begin{equation}
\mathbb{A}_{0}^{+}\rightarrow \frac{\sin (\frac{_{1}}{^{2}}\theta )}{\sqrt{2}%
}(e^{-i\phi }\mathbb{A}_{1}^{+}\mathbb{A}_{3}^{+}+\mathbb{A}_{2}^{+}\mathbb{A%
}_{4}^{+})+\frac{\cos (\frac{_{1}}{^{2}}\theta )}{\sqrt{2}}(e^{-i\phi }%
\mathbb{A}_{1}^{+}\mathbb{A}_{4}^{+}-\mathbb{A}_{2}^{+}\mathbb{A}_{3}^{+}),
\end{equation}%
which then leads to the same predicted probabilities as for the SQM
calculation.

We see here an example of a rank-one state evolving into a rank-two state.
The initial labstate is of rank one at the point of preparation, because
that is the physics of the preparation process. It is only because the
detection equipment is arranged so as to trigger in two separate places that
the final state is legitimately a rank-two state. Without any reference to
the detecting equipment, it is meaningless to talk about the prepared state
being an entangled state of an electron and a positron. Entanglement is not
a fundamental property of a system, but determined by how we observe it.
This is entirely consistent with Bohr's view of the EPR experiment.

\subsection{Two-particle interferometry}

In 1989, Horne, Shimony and Zeilinger discussed an experiment
where an entangled two-photon state passes through a device
consisting of beamsplitters, mirrors, and phase-shifters
\cite{HSZ-1989}. The quantities of interest are the two-particle
coincidence count rates and their dependence on the variable
phase-shift angles, which can be varied at will throughout the
experiment. The conventional representation of the initial state
is
\begin{equation}
|\Psi _{in}\rangle =\frac{1}{\sqrt{2}}\left\{ |\mathbf{k}_{A}\rangle _{1}|%
\mathbf{k}_{C}\rangle _{2}+|\mathbf{k}_{D}\rangle _{1}|\mathbf{k}_{B}\rangle
_{2}\right\} ,
\end{equation}
where the wave vectors $\mathbf{k}_{A}$, $\mathbf{k}_{B}$, $\mathbf{k}_{C}$
and $\mathbf{k}_{D}$ are identified with qubits $\mathcal{Q}_{1}$, $\mathcal{%
Q}_{2}$, $\mathcal{Q}_{3}$ and $\mathcal{Q}_{4}$ respectively, and
subscripts $1$ and $2$ refer to the two particles involved.

The QDN account of this experiment\cite{1} gives the dynamical rule
\begin{equation}
\mathbb{A}_{0}^{+}\rightarrow \frac{1}{\sqrt{2}}\left\{ \mathbb{A}_{1}^{+}%
\mathbb{A}_{3}^{+}+e^{i\theta }\mathbb{A}_{2}^{+}\mathbb{A}_{4}^{+}\right\} ,
\end{equation}%
where the angle $\theta $ depends on the detailed placement of
the various pieces of equipment \cite{HSZ-1989} at the start of
the network, which is the point at which the two-photonic
properties start to manifest themselves. The detailed QDN\
calculation \cite{1} shows how this entangled rank-two state
evolves, in agreement with the calculation of Horne, Shimony and Zeilinger
\cite{HSZ-1989} for the two-particle coincidence probabilities.

\subsection{Other scenarios involving higher rank labstates}

Obvious candidate experiments involving higher rank labstates which remain
to be discussed via the QDN approach are $i)$ interference of photons from
different sources, $ii)$ teleportation, $iii)$ experiments where a sequence
of wave-pulses is set up moving towards target detectors, and $iv)$ large
scale aggregates of quantum optics modules involving beamsplitters, mirrors
and phase-shift devices.

\section{Critique and concluding remarks}

Many interesting physical ideas remain to be explored in QDN. Apart from the
technical challenge of applying the QDN description to ever more complex
experimental situations, there remains the question which dominates modern
theory, viz, the relationship between quantum mechanics and general
relativity. This continuation of the Bohr-Einstein debate arises because of
an almost universal belief in the system concept. By avoiding this concept
directly, the QDN approach bypasses this debate and all the conceptual
problems with it.

However, the picture is not yet complete. It is one thing to write down a
QDN expression analogous to a path integral for a scattering amplitude; it
is another to understand why the various qubits in the network were involved
in the first place. Something is still missing. Until we can understand why
a particular network has the structure that it has, we cannot claim that
quantum mechanics is a complete description of reality. Although we believe
that Bohr won his debate, Einstein was right to doubt the completeness of
SQM. Feynman, who disdained philosophy, understood that this point is a
legitimate one for physicists to think about and this has motivated us to
keep looking into this issue.

Central to this issue is the status of space. We do not believe
it is something which exists in its own right. From the QDN
point of view, space is a potential for observation and no more
than that. That is why in experiments such as the SG experiment,
we need focus only on three qubits. If we wanted to go much
further and discuss for example black hole physics, we would
have to be much more careful in the construction of our
quantized detector network. It would involve vastly more qubits.
In particular, we would have to understand much more about the
dynamical relationship between the evolution of the labstate and
the evolution of the network. Such an understanding would give us
the quantum version of the classical understanding of the
interaction between matter and spacetime which is encoded in
Einstein's equation $G^{\mu \nu }=\kappa T^{\mu \nu }$ in
general relativity.

\section*{Acknowledgements}

I am indebted to Lino Buccheri, Mark Stuckey, Metod Saniga, Jon Eakins, and
my students Jason Ridgway-Taylor and Thomas Marlow for many discussions. I
am also very grateful to Michel Planat for his invaluable support at a
critical time.

\end{document}